\documentclass[aps,prl,showpacs,twocolumn, bibliography]{revtex4-2}

\usepackage{graphicx}
\usepackage{dcolumn}
\usepackage{bm}
\usepackage{color}
\usepackage{amsmath}
\usepackage{amssymb}
\usepackage{comment}
\usepackage{xcolor}
\newtheorem{theorem}{Theorem}
\usepackage[normalem]{ulem}

\DeclareRobustCommand{\erase}{\bgroup\markoverwith{\textcolor{red}{\rule[.5ex]{2pt}{0.4pt}}}\ULon}
 
\begin{document}

%\preprint{APS/123-QED}

\title{Sequential minimum optimization algorithm with small sample size estimators}% Force line breaks with \\

\author{Wojciech Roga$^{1}$}
 \email{wojciech.roga@keio.jp}

\author{Takafumi Ono$^{2,3}$}
 \email{ono.takafumi@kagawa-u.ac.jp}

\author{Masahiro Takeoka$^{1,4}$}
 \email{takeoka@elec.keio.ac.jp}

\affiliation{%
~\\
$^{1}$Department of Electronics and Electrical Engineering, Keio University, 3-14-1 Hiyoshi, Kohoku, Yokohama, Kanagawa 223-8522, Japan} 

\affiliation{%
$^{2}$Program in Advanced Materials Science
Faculty of Engineering and Design,
Kagawa University,
2217-20 Hayashi-cho, Takamatsu, Kagawa
761-0396, Japan
}

\affiliation{%
$^{3}$JST, PRESTO, 4-1-8 Honcho, Kawaguchi, Saitama 332-0012, Japan
} 

\affiliation{%
$^{4}$Advanced ICT Research Institute, National Institute of Information and Communications Technology (NICT), 4-2-1 Nukuikita, Koganei, Tokyo 184-8795, Japan
}

%\affiliation{%
%$^{4}$Advanced ICT Research Institute, National Institute of Information and Communications Technology (NICT), Koganei, Tokyo 184-8795, Japan
%}

%\affiliation{%
%$^{5}$Advanced ICT Research Institute, National Institute of Information and Communications Technology, 588-2 Iwaoka, Nishi, Kobe 651-2492, Japan
%}

\date{\today}

\begin{abstract}
Sequential minimum optimization is a machine-learning global search training algorithm. It is applicable when the functional dependence of the cost function on a tunable parameter given the other parameters can be cheaply determined. This assumption is satisfied by quantum circuits built of known gates. We apply it to photonics circuits where the additional challenge appears: low frequency of coincidence events lowers the speed of the algorithm. We propose to modify the algorithm such that small sample size estimators are enough to successfully run the machine learning task. We demonstrate the effectiveness of the modified algorithm applying it to a quantum optics classifier with data reuploading.
\end{abstract}

\maketitle

\section{Motivation}

There are a variety of proposals how to use quantum devices in machine learning \cite{Mitarai2018,Lloyd2020,Schuld2020,Schuld2021,Chen2021,Cerezo2021}. The research is motivated by the expectation that quantum features can bring benefits for the performance. Indeed, many algorithms use linear algebra modules, which where shown to be solved faster under some conditions when applied on a quantum device \cite{Lewenstein1991,Harrow2009,Lloyd2013,Rebentrost2014,Paparo2014,Georgescu2014,Biamonte2017,Harris2017,Shen2017,Steinbrecher2019,Yu2019,Cong2019,Shastri2021,Zhao2021}. Also Grover search algorithm if applicable can speed up the algorithm \cite{Grover1997}. Or, sampling from a programmable quantum device can be faster than calculating the expectation values of certain observables \cite{Aronson2011,Tillmann2013,Lund2014,Gard2015,Huh2015,Wang2017,Sparrow2018,Quesada2019}. Moreover, the access to entangled states allow one to produce different statistical dependencies than only classical states on a fixed processors as the entangled states occupy a significant volume of all quantum states of a given system \cite{Zyczkowski,Perez2020}.

Among the supervised quantum machine learning algorithms \cite{Havlicek2019,Benedetti_2019,PhysRevA.102.032420,PhysRevA.103.032430,PRXQuantum.2.040316,PhysRevLett.126.190505,Uvarov_2021} we find hybrid quantum-classical ones with classically programmable circuits. Their parameters are trained by classical algorithms such as gradient descent, and the cost function is constructed based on the expectation values of observables that are measured. The speed at which the training algorithms find the global minimum of the cost function is part of the efficiency of the total algorithm.   

There are local search algorithms such as gradient descent, in which the search starts from a chosen starting point and explores the cost function landscape further from that point gradually. Also global search algorithms recognise and explore some global features of the landscape.

One of the global search algorithms proposed for the quantum circuits recently is the so-called sequential minimum optimization \cite{Nakanishi2020} (SMO). It is based on the idea, that from a small number of measurements, one can infer the functional dependence of the cost function along one parameter that typically is a known quantum gate or an optical element like a phase shifter \cite{Matthews2009}. Knowing this, we can in one step update this parameter with the value that corresponds to the minimum of the cost function along this variable. This procedure can be applied sequentially to all parameters of the circuit converging in a few rounds to the minimum of the cost function landscape, which is possibly far from the starting point. This algorithm, unlike local search algorithms, is not affected by barren plateaus and does not rely on additional parameters such as the size of the step. In \cite{Nakanishi2020} where the authors introduced this procedure for the qubit quantum circuits, the convergence was also discussed. 

In \cite{Ono2022}, sequential minimum optimization was used to an integrated photonics based quantum classifier in an experiment that showed computational abilities of two-mode quantum circuit. This paper followed \cite{Perez2020}, in which the authors first showed that a sufficiently long single qubit quantum circuit satisfies the universal approximation theorem \cite{Cyrbenko1989}, known from the theory of a single-layer neural networks, and is able to classify data points divided by an arbitrary border function. The authors performed simulations of several problems using gradient descent training.    
 
However in the photonic circuit experiment with two photons performed in \cite{Ono2022} the sequential minimal optimization was preferred as the time of measuring expectation values of coincident photons is longer. Coincidence counts in this experiment allowed for reducing decoherence errors. However, this type of experiment is strongly affected by losses which induce a long time of gathering enough data in postselection. 

%The cost function was a sumed over the training points the squared error function of probabilities of particular coincidence $p$ which is a function of circuit parameters $\phi$ and the binary classification variable $y$

The experimental run-time needed to train the circuit with the SMO algorithms is proportional to:
\begin{eqnarray}
t &=& \# rounds * \# circuit\_parameters * \# training\_points\nonumber \\
&&* \# function\_coefficients *\# estimator\_measurements,  \nonumber
\end{eqnarray}
where the last term is the number of measurements, we need to estimate the expectation value of the observable measured in the experiment and used in the cost function. The number of function coefficients is the number of values for a given training point needed to specify the functional dependence of the cost function on one parameter of the circuit in the SMO algorithm. The number of rounds is how many times the algorithm updates all parameters. The number of circuit parameters and the number of training points are clear.

The number of rounds can be kept small due to the efficiency of the sequential minimal optimization algorithm \cite{Nakanishi2020}. In the experiment reported in \cite{Ono2022}, it was equal to 10. The number of function parameters, as we will explain later discussing details of the algorithm, is $O(n)$, where $n$ is the number of photons in the circuit. The number of detector clicks to estimate the expectation value of an observable, in principle, can be arbitrary large and in experiment with coincidence measurements can dominate the total time of the training process.

The goal of this paper is to modify the sequential minimal optimization algorithm such that the expectation value of the observable is replaced there by the estimator of this value based on a small sample size. If the number of coincidence events used to estimate a given expectation value of the measured observation was around 200, our modification can reduce the experiment run-time even by 100 times. With this reduction the weak sources generating for instance, entangled states can still be efficiently used in machine learning experiments with many training data points.

This approach is inspired by \cite{Sweke2020}, where the authors discuss in detail the problem of replacing expectation values by unbiased estimators for the gradient descent type training. Then, estimation of the first and sometimes second derivatives of the cost functions at given points are needed. The authors show that the training based on the estimators works well and the computational time related to the measurements can be significantly reduced. The driving idea of the method is the observation that since the cost function is an unbiased estimator of the sum of expectation values of independent variables, it can be replaced by the sum of unbiased estimators of the variables. The error in using estimators from a finite number of measurements depends on the number of measurements and the number of training points. Therefore, if the number of training points is sufficiently large, the number of measurements used for the estimators of the expectation value can be kept small.

In this manuscript, we discuss in detail the idea of \cite{Sweke2020} adapted to the sequential minimal optimization. As \cite{Sweke2020} covers gradient descent-type training, we add something novel to the development of the SMO method. 

The main result is formulated as Theorem 1, it is proven and supported by the simulation of a 2-mode photonic optical classifier with a NOON state as the input. Although the main message of our paper does not require the NOON state, and could be shown with other fixed numbers of photon states, we decided to use the NOON state as a tribute to Jonathan Dowling \cite{Dowling2003}, our friend and collaborator. There are other advantages from using entangled states in this kind of experiments, but it is not our goal to discuss them here.

The paper is organized as follows. First, %in Sec. \ref{sec:SMO}, 
we give details of the sequential minimal optimization for a photonic circuit. This method follows the analogous algorithm derived earlier in \cite{Nakanishi2020} for qubit circuits however our notation and conventions are slightly different. Then, %In Sec. \ref{sec:result}, 
we present the main result of our paper in the form of a Theorem; we give a detailed proof and discuss the convergence of the estimator of the cost function to the real cost function when the number of training points increases. In the following section % Sec. \ref{sec:simulation} 
we present numerical simulator of a photonic classifier. We summarize our work with conclusions and an outlook.

\section{Sequential minimal optimization in a photonic circuit}\label{sec:SMO}

\begin{figure}
    \centering
    \includegraphics[scale = 0.4]{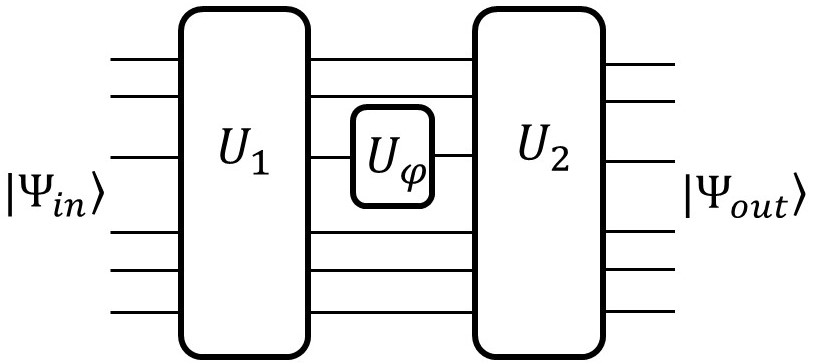}
    \caption{General passive photonic circuit with the input state $|\Psi_{in}\rangle$ of a fixed number of photons. We distinguish a phase shifter depending on a parameter $\varphi$. We want to know the dependence of the probability of $|\Psi_{out}\rangle$ at the end of the circuit on $\varphi$. Here, unitary transformations $U_1$ and $U_2$ can be arbitrary photon number preserving transformations.}
    \label{fig:circuit}
\end{figure}

We consider a general passive optical circuit. Without losing generality, we can compose the circuit with phase-shifters and 50-50 beam-splitters applying for example, the so-called Reck configuration \cite{Reck1994,Clements2016}. Therefore, we can assume that the parameters of the circuit are phases of the phase-shifters. However, the method we discuss below can be easily adapted to circuits with any passive optical elements single- or multi-mode.

We assume that an $n$-photon state $|\Psi_{in}\rangle$ is put as the input and an $n$-photon state $|\Phi_{out}\rangle$ is measured. They can be arbitrary. We assume that a given data point, possibly repeatedly (data-reuploading), is encoded in some phases. 

We consider the cost function to be a function of the probability of measuring an $n$-photon state $|\Psi_{out}\rangle$, trainable phases, and data. For a training point indexed by $i$, let us write the probability of measuring $|\Psi_{out}\rangle$ distinguishing one of the trainable variables $\varphi$ to as in figure \ref{fig:circuit},
\begin{equation}
p^i(\varphi) = |\langle \Psi_{out}|U_2^\dagger U_\varphi U_1|\Psi_{in}\rangle|^2 .
\label{prob}
\end{equation}
Here, $U_1$ and $U_2$ are unitaries before and after a phase-shifter $U_\varphi$ parameterized by $\varphi$. The idea is to update each parameter $\varphi$ to the value where the cost function expressed as the function of this parameter achieves its minimum. To do that we need to derive an explicit formula for $p^i(\varphi)$ for each $\varphi$ in each step of the training. Therefore, assume we are interested in the functional dependence of $p^i$ on phase $\varphi$ of a phase-shifter in mode $m$. Let us expand $U_1|\Psi_{in}\rangle$ and $U_2|\Psi_{out}\rangle$ in the Fock basis writing mode $m$ as the first one
\begin{eqnarray}
U_1|\Psi_{in}\rangle &=& \alpha_0|0\rangle_m|\xi_0\rangle + \alpha_1|1\rangle_m|\xi_1\rangle + ...+\alpha_n|n\rangle|\xi_n\rangle,\nonumber\\
U_2|\Psi_{out}\rangle &=& \beta_0|0\rangle_m|\xi'_0\rangle + \beta_1|1\rangle_m|\xi'_1\rangle + ...+\beta_n|n\rangle|\xi'_n\rangle.\nonumber
\end{eqnarray}
Acting by the phase-shifter on $U_1|\Psi_{in}\rangle$ we get
\begin{eqnarray}
U_{\phi}U_1|\Psi_{in}\rangle &=& \alpha_0|0\rangle_m|\xi_0\rangle + \alpha_1e^{i\varphi}|1\rangle_m|\xi_1\rangle  \nonumber\\
&+&\alpha_2e^{2i\varphi}|2\rangle_m|\xi_2\rangle+...+\alpha_ne^{ni\varphi}|n\rangle|\xi_n\rangle\nonumber.    
\end{eqnarray}

%In this basis 
%\begin{equation}
%    U_\phi = \begin{pmatrix}e^{2i\phi}&0&0\\0&e^{i\phi}&0\\0&0&1\end{pmatrix}.
%\end{equation}
Hence, it easily follows from (\ref{prob}) that
\begin{eqnarray}
p^i(\varphi) &=& A_0 + A_1\cos(\varphi) + A_2 \sin(\varphi)\nonumber\\ 
&+& A_3 \cos(2\varphi) + A_4 \sin(2\varphi)+...,
\label{proba}
\end{eqnarray}
%where
%\begin{eqnarray}
%A_0 &=& |\alpha_1|^2|\beta_1|^2+|\alpha_2|^2|\beta_2|^2+|\alpha_3|^2|\beta_3|^2+...,\label{Azero}\\
%A_1 &=& 2\Im(\bar\alpha_1\alpha_2\beta_1\bar\beta_2) + 2\Im(\bar\beta_1\beta_2\gamma_1\bar\gamma_2),\\
%A_2 &=&-2\Re(\bar\alpha_1\alpha_2\beta_1\bar\beta_2)-2\Re(\bar\beta_1\beta_2\gamma_1\bar\gamma_2),\\
%A_3 &=&2\Im(\bar\alpha_1\alpha_2\gamma_1\bar\gamma_2),\\ 
%A_4 &=&2\Re(\bar\alpha_1\alpha_2\gamma_1\bar\gamma_2).\label{A4}
%\end{eqnarray}
If we knew $A_0,...,A_{2n+1}$, we would have $p^i(\varphi)$. Values $A_j$ can be obtained by solving the set of $2n+1$ equations given by (\ref{proba}) measured for $2n+1$ different $\varphi$. For simplicity, let us assume $n=2$. Then, we can figure out the values of $A_j$ from measuring probabilities denoted here by $r_j=p^i(\varphi_j)$ for $5$ different values of the phase (generalization of this procedure is straightforward)
\begin{equation}
\begin{pmatrix}
1&\cos\varphi_0&\sin\varphi_0&\cos2\varphi_0&\sin2\varphi_0\\
1&\cos\varphi_1&\sin\varphi_1&\cos2\varphi_1&\sin2\varphi_1\\
1&\cos\varphi_2&\sin\varphi_2&\cos2\varphi_2&\sin2\varphi_2\\
1&\cos\varphi_3&\sin\varphi_3&\cos2\varphi_3&\sin2\varphi_3\\
1&\cos\varphi_4&\sin\varphi_4&\cos2\varphi_4&\sin2\varphi_4
\end{pmatrix}  
\begin{pmatrix}
A_0\\A_1\\A_2\\A_3\\A_4
\end{pmatrix}=
\begin{pmatrix}
r_0\\r_1\\r_2\\r_3\\r_4
\end{pmatrix}.
\label{cosmatrix}
\end{equation}
Values $r_k$ are the probabilities measured by the same circuit with 5 chosen phases $\varphi_k$. In particular, if one choses 
\begin{eqnarray}
\varphi_0 = 0,\quad \varphi_1 = \frac{2\pi}{5},\quad \varphi_2 = -\frac{2\pi}{5},\quad
\varphi_3 = \frac{4\pi}{5},\quad
\varphi_4 = -\frac{4\pi}{5}\nonumber,
\end{eqnarray}
equation (\ref{cosmatrix}) takes the form of the easily invertible Fourier transform
\begin{equation}
F  
\begin{pmatrix}
(A_3 - iA_4)/2\\(A_1-iA_2)/2\\A_0\\(A_1 + iA_2)/2\\(A_3+iA_4)/2
\end{pmatrix}=
\begin{pmatrix}
r_3\\r_1\\r_0\\r_2\\r_4
\end{pmatrix},
\label{omegamatrix}
\end{equation}
where the Fourier transform matrix is
\begin{equation}
   F = \begin{pmatrix}
\omega^4&\omega^2&\omega^0&\omega^{-2}&\omega^{-4}\\
\omega^2&\omega^1&\omega^0&\omega^{-1}&\omega^{-2}\\
\omega^0&\omega^0&\omega^0&\omega^0&\omega^0\\
\omega^{-2}&\omega^{-1}&\omega^0&\omega^{1}&\omega^{2}\\
\omega^{-4}&\omega^{-2}&\omega^0&\omega^{2}&\omega^{4}\\
\end{pmatrix},
\label{fourier}
\end{equation}
and $\omega = e^{i\frac{2\pi}{5}}$. Hence
%The inverse of the Fourier matrix is 
%\begin{equation} 
%F^{-1} = \frac{1}{5}F^{\dagger}.
%\end{equation}
%Therefore,
%\begin{equation}
%\begin{pmatrix}
%A_3\\A_1\\A_0\\A_2\\A_4
%\end{pmatrix}= \frac{1}{5}SF^\dagger 
%\begin{pmatrix}
%r_3\\r_1\\r_0\\r_2\\r_4
%\end{pmatrix},
%\end{equation}
%where matrix $S$ is a simple change of variables,
%\begin{equation}
%S = \begin{pmatrix}
%1&0&0&0&1\\
%0&1&0&1&0\\
%0&0&1&0&0\\
%0&i&0&-i&0\\
%i&0&0&0&-i\\
%\end{pmatrix}.
%\end{equation}

\begin{equation}
\begin{pmatrix}
A_3\\A_1\\A_0\\A_2\\A_4
\end{pmatrix}= \frac{2}{5}
\begin{pmatrix}
\cos\frac{8\pi}{5}&\cos\frac{4\pi}{5}&1&\cos\frac{4\pi}{5}&\cos\frac{8\pi}{5}\\
\cos\frac{4\pi}{5}&\cos\frac{2\pi}{5}&1&\cos\frac{2\pi}{5}&\cos\frac{4\pi}{5}\\
0.5&0.5&0.5&0.5&0.5\\
\sin\frac{4\pi}{5}&\sin\frac{2\pi}{5}&0&-\sin\frac{2\pi}{5}&-\sin\frac{4\pi}{5}\\
\sin\frac{8\pi}{5}&\sin\frac{4\pi}{5}&0&-\sin\frac{4\pi}{5}&-\sin\frac{8\pi}{5}
\end{pmatrix}
\begin{pmatrix}
r_3\\r_1\\r_0\\r_2\\r_4
\end{pmatrix}.
\label{formulaforA}
\end{equation}

This shows that to obtain the functional dependence of the probability $p(\varphi)$ in (\ref{proba}) on each adjustable circuit parameter $\varphi$, we need to measure $2n+1$ probabilities $r_j$, where $n$ is the number of photons, with the same circuit but with different values of the parameter under investigation. 

Notice that with the effort of finding $A_j$ for each training point, we can find an explicit formula for the cost function and its minimum for each adjustable parameter. Hence, we can update the parameter with the value in which the cost function achieves the minimum, which is the key idea of SMO. This procedure is applied in the analysis of the following section.

\section{Sequential Minimum Optimization with small sample size estimators}\label{sec:result}

For each training point enumerated by index $i$, for each phase shifter characterized by its $\varphi$, we measure $2n+1$ probabilities $r_j$ for $2n+1$ phases, as in the previous section. From the probabilities we calculate parameters $A_k^i$ that allow us to define the function $p^i(\varphi)$ as in (\ref{proba}) that enters to the cost function

\begin{equation}
    C(\varphi) = \frac{1}{N}\sum_{i=1}^N (p^i(\varphi) - y^i)^2,
\end{equation}
which is still an easy function of $\varphi$ that can be efficiently minimized numerically. The phase of the phase shifter is updated as follows
\begin{equation}
    \varphi\rightarrow {\rm argmin}\ C(\varphi),
\end{equation}
and we move to the next phase shifter. We repeat everything until a stopping criterion is met.

However, in an experiment with photonic circuits to estimate a real single probability $r_j$, we use its estimator $\tilde{r}_j=N_{success}/N_{all}$, which is assumed to be a random variable with a Binomial distribution. The related mean square error is scaled with the sample size $N_{all}$ as $O(1/N_{all})$. To keep the error small, we typically collect data from a few hundreds of coincidence events. The sample size depends on the intensity of the source and time we want to spend for the total experiment. For example, in \cite{Ono2022}, estimating a single probability  $\tilde{r}_j$ took around a second with a weak coherent state source that produced around 3000 two-photon coincidences in this interval. In experiments with multiple photons or with sources of entangled photons, the time of gathering similar statistics would be significantly increased and running the experiment would be inefficient.

However, in what follows, we show that we can successfully run our experiment with strongly smaller sample size estimators, i.e., when $N_{all}$ is small. Indeed, some tasks can be satisfactory solve with as little sample size as $N_{all}=1$ in which case $\tilde{r}_j$ is binary.

Therefore, for each training point indexed by $i$, using the  estimators $\tilde{r}_j^i$, we can calculate the estimators of $A^i_k$ denoted by $\tilde A^i_k$ as in Eq. (\ref{formulaforA}) and define the estimator of  $p^i(\varphi)$ analogous to (\ref{proba}) as follows
\begin{equation}
\tilde p^i(\varphi) = \tilde A^i_0 + \tilde A_1^i \cos(\phi) + \tilde A_2^i \sin(\varphi) + ... 
\label{probb}
\end{equation}
We prove the following theory that shows that for a large number of the training points $N$ a function built from these estimators converges to the cost function built from their expectation values.
\begin{theorem}
Let $N$ be the number of training points and
\begin{equation}
    \tilde C(\varphi) = \frac{1}{N} \sum_i^N \Big(\tilde p^i(\varphi)(\tilde p^i)'(\varphi) - 2\tilde p^i(\varphi)y^i + (y^i)^2\Big),
    \label{costestimator}
\end{equation}
where $\tilde p^i(\varphi)$ and $(\tilde p^i)'(\varphi)$ are functions (\ref{probb}) generated from separate independent measurements for the same training point. For an arbitrary unbiased estimator of the probabilities of experimental outcomes and for sufficiently large $N$
\begin{equation}
{\rm argmin}\ \tilde C(\varphi) \rightarrow {\rm argmin}\  C(\varphi)
\end{equation}
\end{theorem}
{\bf Proof.} Let us expand $\tilde C(\varphi)$
\begin{eqnarray}
\tilde C(\varphi) &=& \frac{1}{N} \sum_i^N \Big[\Big(\tilde A_0^i + \tilde A_1^i\cos(\varphi) + \tilde A_2^i \sin(\varphi) + ...\Big)\nonumber\\
&&\ \ \ \ \ \ \ \ *\Big((\tilde A_0^i)' + (\tilde A_1^i)'\cos(\varphi) + (\tilde A_2^i)' \sin(\varphi) + ...\Big)\nonumber\\
&-& 2 \Big(\tilde A_0^i + \tilde A_1^i\cos(\varphi) + \tilde A_2^i \sin(\varphi) + ...\Big)y^i + (y^i)^2\Big]\nonumber\\
&=& \frac{1}{N} \sum_i^N \tilde A^i_0(\tilde A^i_0)' + \frac{1}{N} \sum_i^N\tilde A^i_0(\tilde A^i_1)'\cos(\varphi) + ... \nonumber\\
&-&2 \frac{1}{N} \sum_i^N\tilde A_0^iy_i -2\frac{1}{N} \sum_i^N \tilde A_1^iy^i \cos(\varphi) +...\nonumber \\
&+&\frac{1}{N} \sum_i^N (y^i)^2\nonumber.
\end{eqnarray}
Let us look first at the term linear in $\tilde A^i_0$. We express it by the estimators of the probabilities $\tilde r^i_k$
\begin{equation}
X = \frac{1}{N} \sum_i^N\tilde A_0^iy_i = \frac{1}{N} \sum_i^N(w^0_0 \tilde r^i_0 + w^0_1 \tilde r^i_1 + ...)y^i, 
\end{equation}
where $w$ are appropriate constants from equation (\ref{formulaforA}).
Notice that as $\tilde r^i_j$ are random variables with expectation values $r^i_j$, also $X$ is a random variable with the expectation value
\begin{equation}
\frac{1}{N} \sum_i^N(w^0_0 r^i_0 + w^0_1 r^i_1 + ...)y^i
\end{equation}
here $r^i_j$ are the values of the real probabilities from the experiment. The standard deviation of $X$ scales with $N$ such as $\frac{\sigma^i}{\sqrt{N}}$, where $(\sigma^i)^2$ is the variance of $\tilde r^i_j$. The expectation value of $X$ is the same as the term computed from the perfect probabilities $r^i_j$, with the standard deviation vanishing with $N$. The same reasoning holds for other linear terms of $\tilde C(\varphi)$. This implies that the linear terms of $\tilde C(\varphi)$ tend to the analogous terms of $C(\varphi)$.

Let us consider now the quadratic term
\begin{eqnarray}
Y&=&\frac{1}{N} \sum_i^N \tilde A^i_0(\tilde A^i_0)'\nonumber\\ 
&=&  \frac{1}{N} \sum_i^N(w^0_0 \tilde r^i_0 + w^0_1 \tilde r^i_1 + ...)(w^0_0 (\tilde r^i_0)' + w^0_1 (\tilde r^i_1)' + ...) \nonumber \\
&=& \frac{1}{N} \sum_i^N w^0_0w^0_0 \tilde r^i_0(\tilde r^i_0)'+...
\end{eqnarray}
For $\tilde r^i_0$ and $(\tilde r^i_0)'$ being independent random variables, $\frac{1}{N} \sum_i^N w^0_0w^0_0 \tilde r^i_0(\tilde r^i_0)'$ is a random variable with the expectation value equal to $(w^0_0)^2(r^i_0)^2$. Similarly, other terms of $Y$. Therefore, the expectation value of $Y$ is $\frac{1}{N} \sum_i^N (w^0_0)^2( r^i_0)^2+...$ that is the same as the analogous term in $C(\varphi)$. By the same argument as previously, the variance of $Y$ vanishes with $N$. The same reasoning can be applied to the other quadratic terms. 

Concluding, the coefficients in front of trigonometric functions in $\tilde C(\varphi)$ tend to the appropriate coefficients of $C(\varphi)$ when $N$ is sufficiently large. QED.

\section{Simulation of photonic classifier with data reuploading}\label{sec:simulation}

\begin{figure}
    \centering
    \includegraphics[scale = 0.35]{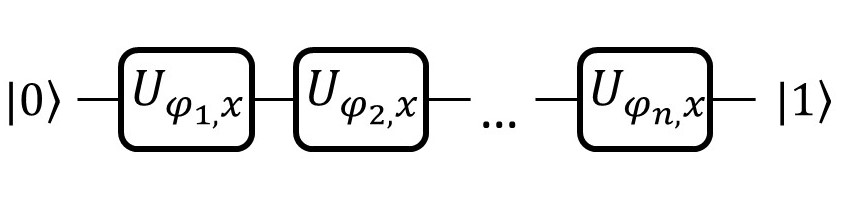}
    \caption{Layers of the single qubit classifier with data re-uploading \cite{Perez2020}.}
    \label{fig:layers}
\end{figure}

As a demonstration of this method, let us simulate a fragment of a simple optical universal classifier. The idea of a universal quantum classifier with a single qubit was developed in \cite{Perez2020}. In this scenario, two-dimensional data points $x$ to be classified are encoded as parameters of a single qubit unitary transformation $U(\varphi,x)$ of the qubit initialized in the ground state $|0\rangle$. This transformation also contains the free parameters $\varphi$, which are tuned during the learning process. %The scheme of the classifier is shown \ref{fig:layers}. %(The layer can be constructed as well as a single unitary transformation the parameters of which depend both on ${\bf x}$ and $\phi$, e.g., $U({\bf x}, \phi)$).
Similarly, more unitary gates can be introduced sequentially with different free parameters, but the same data $x$ as in figure \ref{fig:layers}. This provides the nonlinearity of the expectation values of the output states in the variable encoding the data. After the circuit, the qubit is measured. The classification is based on the results. The free parameters $\varphi$ are adjusted during the training of the system. 

%A single layer can classify only 2D functions which depend on simple squared weighted sums of trigonometric functions of the input. One can obtain more complicated functions if more layers are applied. 

The authors of \cite{Perez2020} provide arguments that in this way, one can arbitrarily approximate any function of the input points achieving a universal classifier. The arguments are based on the analogy to the universality of neural networks.

\begin{figure}
    \centering
    \includegraphics[scale = 0.3]{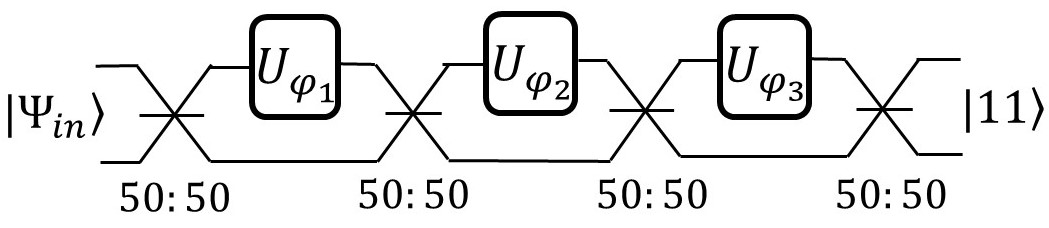}
    \caption{A two mode unitary transformation built of phase-shifters and the Mach-Zender interferometer.}
    \label{fig:single}
\end{figure}

In \cite{Ono2022} the idea of the universal quantum classifier has been implemented on an optical circuit of integrated photonics using the advantages of this platform. The original scheme from \cite{Perez2020} was adopted for bosonic systems. The classification based on the data reuploading scheme has been performed experimentally on an integrated photonic quantum optical circuit. Instead of the qubit, two mode state of two photons was used. In this way losses has been eliminated from the system as the coincidences of two photons where measured.

\begin{figure*}
    \centering
    \includegraphics[scale = 0.4]{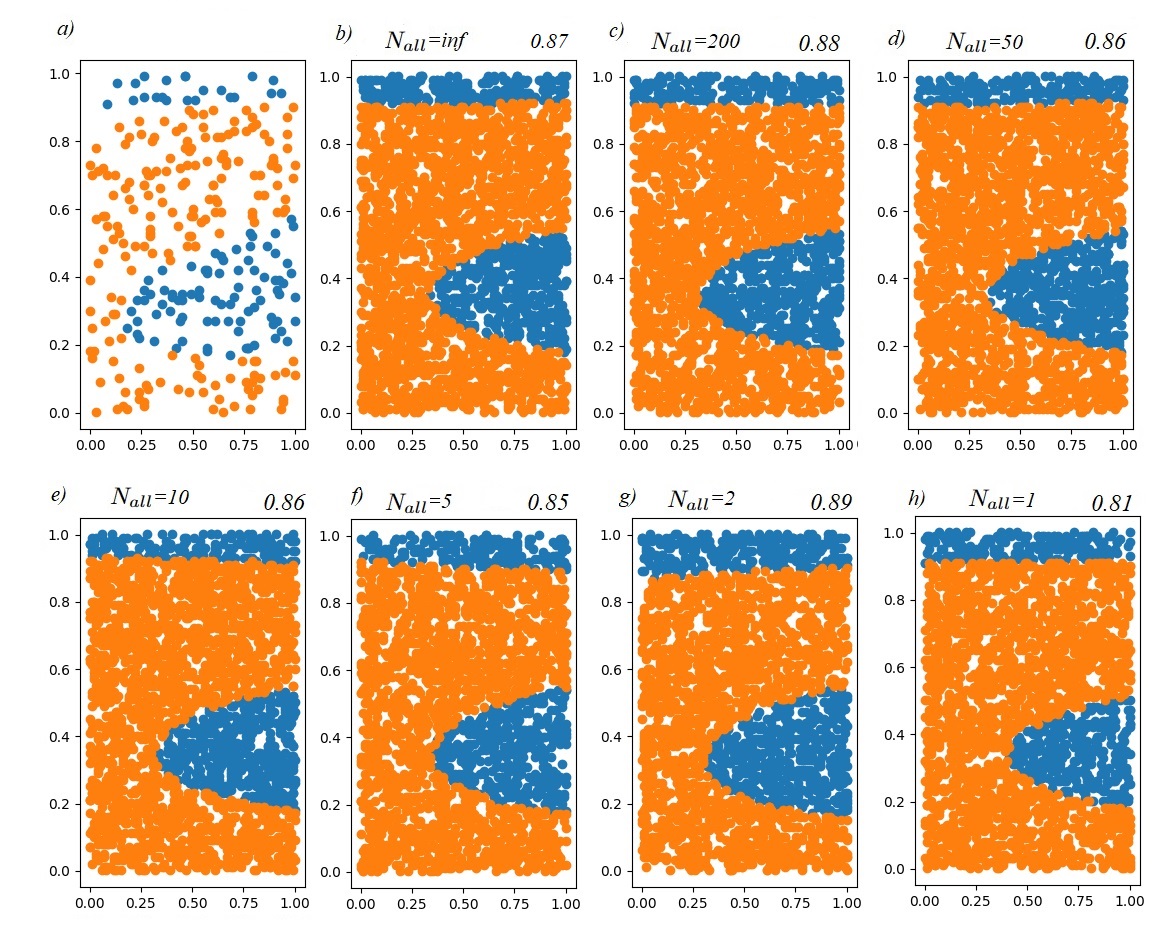}
    \caption{a) Training points in two classes. b)-h) Classifier trained with respectively: probabilites, estimators from $N_{all}=200$, estimators from $N_{all}=50$, estimators from $N_{all}=10$, estimators from $N_{all}=5$, estimators from $N_{all}=2$, estimators from $N_{all}=1$. In the upper right corner of each figure the average success probability is written.}
    \label{fig:grid}
\end{figure*}

We simulate an optical classifier shown in figure \ref{fig:single} to classify two dimensional points in two regions as shown in figure \ref{fig:grid} $a)$. %\ref{fig:trainingtesting}.
%\begin{figure}
%    \centering
%    \includegraphics[scale = 0.45]{Figure_1.png}
%    \caption{Left image: training points with the correct classification. Right image: testing points classified by the classifier.}
%    \label{fig:trainingtesting}
%\end{figure}
%In figure \ref{fig:costfunction} we show the cost function when our minimum-descent algorithm is applied. The classification by the optimal circuit is shown in the right part of figure \ref{fig:trainingtesting}.
The input state is taken as a NOON state
\begin{equation}
    |\Psi_{in}\rangle=\frac{1}{\sqrt{2}}(|20\rangle+|02\rangle)
\end{equation}
(although we would obtain similar results with other input states). The circuit consists of phase-shifters and fixed beam-splitters. The phases are functions of the free parameters adjustable during the training and data. Specifically, one phase $\varphi_j=\phi_{j_1}+\phi_{j_2}x_k$, where $x_k$ is either $x_1$ -- the horizontal coordinate of a 2D data point or $x_2$ -- the vertical coordinate. So, the circuit from figure \ref{fig:single} contains 6 trainable parameters $\phi$. Notice that the same data are re-uploaded repetitively. During the training at the end of the circuit, we measure the estimator of the probability of state $|11\rangle$. We train the circuit minimizing the cost function defined in (\ref{costestimator}). 

After the training we generate new testing points. To test the classifier we calculate the proper probability of $|11\rangle$ and classify a given point as 0 if the probability is smaller than a threshold, otherwise the class is denoted as 1. We check the number of results which are: true positive (TP, classifier indicates 1 for class 1), true negative  (TN, classifier indicates 0 for class 0), false positive  (FP, classifier indicates 1 for class 0), and false negative  (FN, classifier indicates 0 for class 1). The quality of performance is characterised by the average success probability, which is an average of the true positive and true negative ratios:
\begin{equation}
P=\frac{1}{2}TPR+\frac{1}{2}TNR,
\end{equation}
where
\begin{eqnarray}    
    TPR &=& \frac{TP}{TP + FN}, \nonumber\\
    TNR &=& \frac{TN}{TN + FP}.
\end{eqnarray}

\begin{figure}
    \centering
    \includegraphics[scale = 0.65]{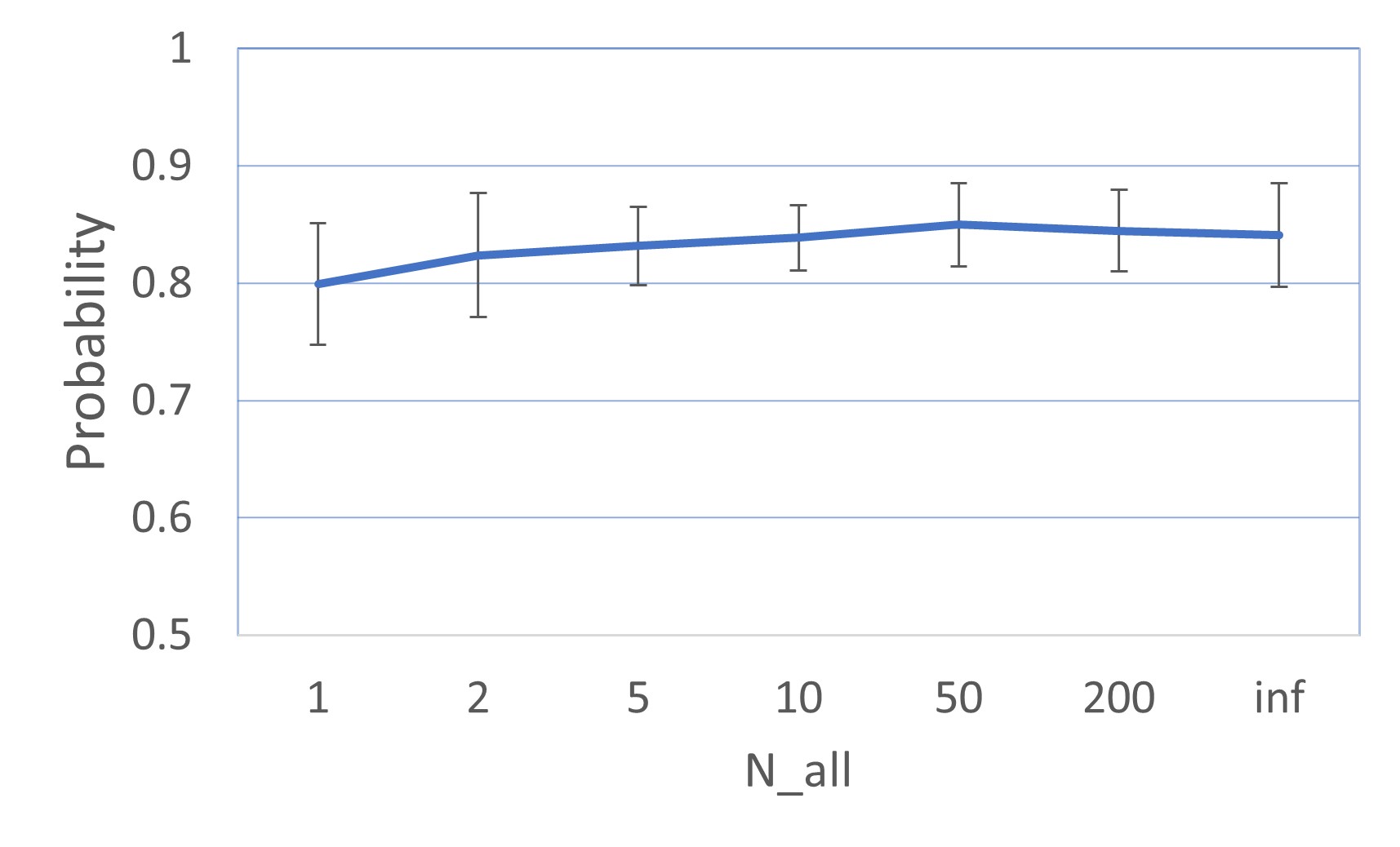}
    \caption{Average of the probability of success as a function of the number of measurements used by the estimator. The average is over different sets of training data. The task and the structure of the circuit are fixed. The mark "inf" denotes the real value of the probability. The error bars show the stadard deviation.}
    \label{fig:aver}
\end{figure}

We repeat the simulation for different estimators of the probability given by the sample size $N_{all}=200, 50, 10, 5, 2, 1$. The number of training points is taken as $N=300$. The training points and the starting point are the same in experiments different $N_{all}$. The results are shown in figures \ref{fig:grid} and \ref{fig:aver}. 

The numerical tests were performed with 300 training data, as shown in figure \ref{fig:grid} $a)$ with 10 rounds of updating all 6 parameters of the circuit. We observe that using an estimator of the probability based on 200 measurements give the same efficiency as the algorithm with the exact probabilities, figure \ref{fig:grid} $b)$. However, we observe that the efficiency only slightly changes when $N_{all}$ is reduced till $N_{all}=1$ when the estimator of a single probability is a binary variable. For all estimators features of the sets are clearly visible. Figure \ref{fig:aver} shows the plot of the probability of success averaged over different series of training data versus the number of measurements $N_{all}$ of the estimators. The averages were calculated by repeating all experiments for 10 different training data sets.

Based on our numerical example, we conclude that an estimator based on 1 measurement is still sufficiently good to recognize points from different classes with a high probability of success. Taking these estimators allows us to reduce the time of the experiment based on 200 measurements per probability, 100 times. 

\section{Discussion}

In the context of machine learning with quantum optical circuits, we study a training algorithm, the so-called sequential minimal optimization \cite{Nakanishi2020, Ono2022} that gives us an update rule for parameters in the optimization of the cost function. This algorithm is a non local search alternative to gradient descent. In the original algorithm, the cost function is constructed from measured estimators of the expectation values of variables. In some types of experiments in which the interesting events are rare, building estimators that well approximate the expectation values takes time that may be prohibitive. Therefore, following the research on gradient descent algorithms with small sample size estimators \cite{Sweke2020}, we develop the theory of the sequential minimum optimization algorithm with small sample size estimators. Our theory allows for using week sources of light to run still efficient supervised machine learning tasks.

Our theory was demonstrated in the photonic circuit experiment but is not restricted to it. It can be used as well in other platforms where the events under interest are rare.

%Similar reasoning as described in the previous sections can be applied to the same circuit with coherent states as input. We notice however that the dependence of the intensity in one output on the parameter $\phi$ is 
%\begin{equation}
%    I = A_0 + A_1 \sin \phi + A_2 \cos\phi,
%\end{equation}
%which is significantly simpler than formula (\ref{proba}). This implies that quantum version provides more complicated dependencies. Which may imply a shorter circuit needed to solve a given problem.\\

\textbf{\emph{Acknowledgements}}
The authors are grateful to Jonathan P. Dowling for his insightful suggestions and  encouragements (and plenty of humor!) to this work. This work was initially started when the authors were working at National Institute of Information and Communications Technology (NICT). Jon was a frequent visitor to NICT and initiated many projects and collaborations, where this work is a part of them. He will be immenselly missed. 
This work was supported by JST PRESTO Grant No.~JPMJPR1864, The Murata Science Foundation, JST CREST Grant No.~JPMJCR1772, and JST COI-NEXT Grant No.~JPMJPF2221.

\bibliography{references}

\end{document}